\documentclass[preprint]{aastex}

\def\teff{T$_{\rm{eff}}\,$}
\def\teffs{T$_{\rm{eff}}s\,$}

\def\mj{$M_{\rm J}\,$}

\def\Dwa{$\,$\uppercase\expandafter{\romannumeral5}$\,$}
\def\mic{$\mu$m$\,$}

\def\sless{\lower2pt\hbox{$\buildrel {\scriptstyle <}
   \over {\scriptstyle\sim}$}}
\def\sgreat{\lower2pt\hbox{$\buildrel {\scriptstyle >}
   \over {\scriptstyle\sim}$}}
\def\aa{Astron. Astrophys.\ }

\begin{document}

\title{Beyond the T Dwarfs: Theoretical Spectra, Colors, and Detectability of the Coolest Brown Dwarfs}

\author{Adam Burrows\altaffilmark{1}, David Sudarsky\altaffilmark{1}, \&  Jonathan I. Lunine\altaffilmark{2}}

\altaffiltext{1}{Department of Astronomy and Steward Observatory, 
                 The University of Arizona, Tucson, AZ \ 85721}
\altaffiltext{2}{Lunar and Planetary Laboratory, The University of Arizona, Tucson, AZ \ 85721}

\begin{abstract}

We explore the spectral and atmospheric properties of brown dwarfs cooler than 
the latest known T dwarfs.  Our focus is on the yet-to-be-discovered free-floating brown dwarfs in the 
\teff range from $\sim$800 K to $\sim$130 K and with masses from 25 to 1 \mj.
This study is in anticipation of the new characterization capabilities 
enabled by the launch of SIRTF and the eventual launch of JWST.  In addition, it is in support of the
continuing ground-based searches for the coolest substellar objects.  
We provide spectra from $\sim$0.4 \mic to 30 \mic, highlight the evolution and mass dependence
of the dominant H$_2$O, CH$_4$, and NH$_3$ molecular bands, consider the formation
and effects of water-ice clouds, and compare our theoretical flux densities with 
the putative sensitivities of the instruments on board SIRTF and JWST.  The latter
can be used to determine the detection ranges from space of cool brown dwarfs.
In the process, we determine the reversal point of the blueward trend in the near-infrared colors
with decreasing \teff (a prominent feature of the hotter T dwarf family),   
the \teffs at which water and ammonia clouds appear, the strengths of gas-phase ammonia 
and methane bands, the masses and ages of the objects for which the 
neutral alkali metal lines (signatures of L and T dwarfs) are muted,
and the increasing role as \teff decreases of the mid-infrared fluxes longward of 4 \mic.
These changes suggest physical reasons to expect the emergence of at least one
new stellar class beyond the T dwarfs.
Furthermore, studies in the mid-infrared could assume a new, perhaps transformational,
importance in the understanding of the coolest brown dwarfs.  
Our spectral models populate, with cooler brown dwarfs 
having progressively more planet-like features, the theoretical gap between 
the known T dwarfs and the known giant planets.  Such objects likely inhabit the  
galaxy, but their numbers are as yet unknown.

\end{abstract}

\section{Introduction}
\label{intro}

The discovery of Gliese 229B (Oppenheimer et al. 1995) and 
the successes of the 2MASS (Reid 1994; Stiening, Skrutskie, and Capps 
1995), Sloan (Strauss et al. 1999), and DENIS (Delfosse et al. 1997) 
surveys have collectively opened up a new chapter in stellar astronomy.
The L and T dwarfs (Kirkpatrick et al. 1999,2000; Mart{\'{\i}}n et al. 1999; Burgasser et al. 1999, 2000a,b,c)
that have thereby been discovered and characterized comprise the first new  
``stellar" types to be added to the stellar zoo in nearly 100 years.   
The lower edge of the solar-metallicity main sequence is an L dwarf
not an M dwarf, with a \teff near 1700 Kelvin (K),
and more than 200 L dwarfs spanning a \teff range
from $\sim$2200 K to $\sim$1300 K are now inventoried.  The coolest L dwarfs are also brown dwarfs,
objects too light ($\sless 0.074$ M$_{\odot}$) to ignite hydrogen stably on the main sequence (Burrows et al. 2001).  Similarly, 
to date approximately 40 T dwarfs have been discovered spanning the \teff
range from $\sim$1200 K to $\sim$750 K.  These are all brown dwarfs
and are the coldest ``stars" currently known.  

However, the edge of the ``stellar" mass
function in the field, in the solar neighborhood, or in star clusters has not yet been reached and it
is strongly suspected that in the wide mass and \teff gap between the currently known T dwarfs
and Jovian-like planets there resides a population of very cool ($ 150\, {\rm K} < T_{\rm eff} < 800$ K)
brown dwarfs.  Such objects could be too dim in the optical and near-infrared to have been 
seen with current technology, but might be discovered in the not-too-distant
future by the NGSS/WISE infrared space survey (Wright et al. 2001), SIRTF 
(Space InfraRed Telescope Facility; Werner and Fanson 1995), and/or 
JWST (James Webb Space Telescope; Mather and Stockman 2000).
In this paper, we calculate the spectra and colors of such a population
in order to provide a theoretical underpinning for the future study
of these coolest of brown dwarfs.  Dwelling as they do at \teffs beyond those of the currently-known 
T dwarfs, these ``stars" emit strongly in the near- and mid-infrared.  
Consequently, we highlight their fluxes from 1 to 30 microns and compare these
fluxes with the putative sensitivities of instruments on SIRTF and 
planned instruments on JWST.  We include the effects of water clouds that form 
in the coolest of these objects.  The presence of clouds of any sort emphasizes 
the kinship of this transitional class with solar system planets, in which clouds
play a prominent role.  (Note, however, that on Jupiter itself water clouds 
are too deep below the ammonia cloud layer to have been unambiguously detected.) 

Since we focus on isolated free-floaters or wide binary brown dwarfs, we do not include
external irradiation by companions.  The \teffs of this model set ($\le 800$ K) are such that silicate and iron
clouds are deeply buried.  Hence, unlike for L dwarfs and 
early T dwarfs (Marley et al. 2002; Burrows et al. 2002), 
the effect of such refractory clouds on emergent spectra
can be ignored.  In \S\ref{approach}, we discuss our numerical
approaches and inputs.  We go on in \S\ref{models} to describe our mass-age model set
and our use of the Burrows et al. (1997) evolutionary calculations to provide the 
mapping between (\teff, gravity [$g$]) and (mass, age). In \S\ref{profiles}, we present a representative
sample of derived atmospheric temperature(T)/pressure(P) profiles and their systematics.
This leads in \S\ref{sense} to a short discussion of the SIRTF and JWST point-source sensitivities.
Section \ref{spectra} concerns the derived spectra and is the central section of the paper.
In it, we discuss prominent spectral features from the optical to 30 microns, trends
as a function of age, \teff, and mass, diagnostics of particular atmospheric constituents,
and detectability with instruments on SIRTF and JWST.  We find that these
platforms can in principle detect brown dwarfs cooler than the current T dwarfs out to large distances.  We also
explore the evolution of $J-K$ and its eventual return to the ``red" 
(Marley et al. 2002; Stephens, Marley, and Noll 2001), reversing the blueward trend with decreasing \teff that roughly characterizes the known
T dwarfs.  Furthermore, we make suggestions for filter sets that may
optimize their study with NIRCam on JWST.  Finally, we present physical reasons for anticipating the emergence
of a new stellar type beyond the T dwarfs.  In \S\ref{conclusion}, we summarize what we have determined about this
coolest-dwarf family and the potential for their 
detection.

\section{Numerical Tools and Assumptions}
\label{approach}

To calculate model atmospheres of cool brown dwarfs requires 1) a method 
to solve the radiative transfer, radiative equilibrium, and hydrostatic
equilibrium equations, 2) a convective algorithm, 3) an equation of state that also provides
the molecular and atomic compositions, 4) a method to model clouds
that may form, and 5) an extensive opacity
database for the constituents that arise in low-temperature,
high-pressure atmospheres.  The computer program we use to solve the atmosphere
and spectrum problem in a fully self-consistent fashion is an updated version
of the planar code TLUSTY (Hubeny 1988; Hubeny \& Lanz 1995), which uses a hybrid of Complete
Linearization and Accelerated Lambda Iteration (Hubeny 1992).  To handle convection, 
we use mixing-length theory (with a mixing length equal to one pressure-scale height). The 
equation of state we use to find the P/T/density($\rho$) relation is that of Saumon, 
Chabrier, and Van Horn (1995) and the molecular compositions are calculated
using a significantly updated version of the code SOLGASMIX (Burrows and Sharp 1999).
The latter incorporates a rainout algorithm for refractory silicates and iron (Burrows et al. 2001).
The most important molecules are H$_2$, H$_2$O, CH$_4$, CO, N$_2$, and NH$_3$
and the most important atoms are Na and K.  

We determine when water condenses 
by comparing the water ice condensation curve (the total pressure at which
the partial pressure of water is at saturation) with the object's T/P profile. For pressures
lower than that near the associated intercept, we deplete the vapor phase through the expected 
rainout and embed an absorbing/scattering water-ice cloud with a thickness 
of one pressure-scale-height in the region above.  Note that the total gas pressures at which
the partial pressure of water is at the triple-point pressure of water
are generally higher than the intercept pressures we find.  Hence, the water gas to water ice 
(solid) transition is the more relevant.  Note also that the optical properties
of water ice and water droplets are not very different.  The ice particles are assumed to be spherical and their
modal particle radii are derived using the theory of Cooper et al. (2003).  They vary in size
from $\sim$20 \mic (higher-$g$/lower-\teff) to $\sim$150 \mic (lower-$g$/higher-\teff)
and we assume that the particle size is independent of altitude.  A canonical super-saturation factor
(Cooper et al. 2003; Ackermann and Marley 2001) of 0.01 (1.0\%) is used.  Curiously, with such large particles and such
a small super-saturation, the absorptive opacity of our baseline water-ice clouds, when they do form, is not  
large.  In fact, the consequences for the emergent spectrum of the associated drying of the upper atmosphere,
and the corresponding diminution of the water vapor abundance there, are comparable to
the effects on the spectrum of the clouds themselves.  Without an external flux source, and the scattering
of that flux back into space by clouds, water-ice clouds seem to have only a secondary influence on the 
spectra of the coolest isolated brown dwarfs.  

We use the constantly-updated opacity database described in 
Burrows et al. (1997,2001,2002). This includes Rayleigh scattering,
Collision-Induced Absorption (CIA) for H$_2$ (Borysow and Frommhold 1990; 
Borysow, J\o{rgensen}, and Zheng 1997), and T/P-dependent absorptive 
opacities from 0.3 \mic to 300 \mic for H$_2$O, CH$_4$, CO, and  NH$_3$. 
The opacities of the alkali metal atoms are taken from Burrows, Marley, and Sharp (2000),
which are similar in the line cores and near wings to those found in Burrows and Volobuyev (2003).
The opacities are tabulated in T/$\rho$/frequency space using the abundances derived for a solar-metallicity
elemental abundance pattern (Anders and Grevesse 1989; Grevesse and Sauval 1998; Allende-Prieto, Lambert, and Asplund 2002).
During the TLUSTY iterations, the opacity at any thermodynamic point 
and for any wavelength is obtained by interpolation.  The absorptive opacities for 
the ice particles are derived using Mie theory with the frequency-dependent spectrum 
of the complex index of refraction of water ice.  
Ammonia clouds form in the upper atmospheres of the coldest exemplars of 
the late brown dwarf family (\teff $\le$ 160 K; \S\ref{profiles}).
Nevertheless, since the scattering of incident radiation that gives them their true importance in the Jovian context
is absent, we ignore them here.

\section{Model Set}
\label{models}

We have chosen for this study a set of models with the masses and ages
given in Table 1.  Also shown in Table 1 are the corresponding gravities and \teffs.
These models span an effective temperature range from $\sim$800 K to $\sim$150 K
that allows us to probe the realm between the known T dwarfs and the known Jovian planets.
To establish the mapping between mass/age pairs and the \teff/$g$ pairs that are needed
for atmospheric calculations, we use the evolutionary models of Burrows et al. (1997).  While
this procedure does not ensure that the atmospheres we calculate 
are fully consisitent with those evolutionary tracks,
the errors are not large.  

\begin{figure}
\plotone{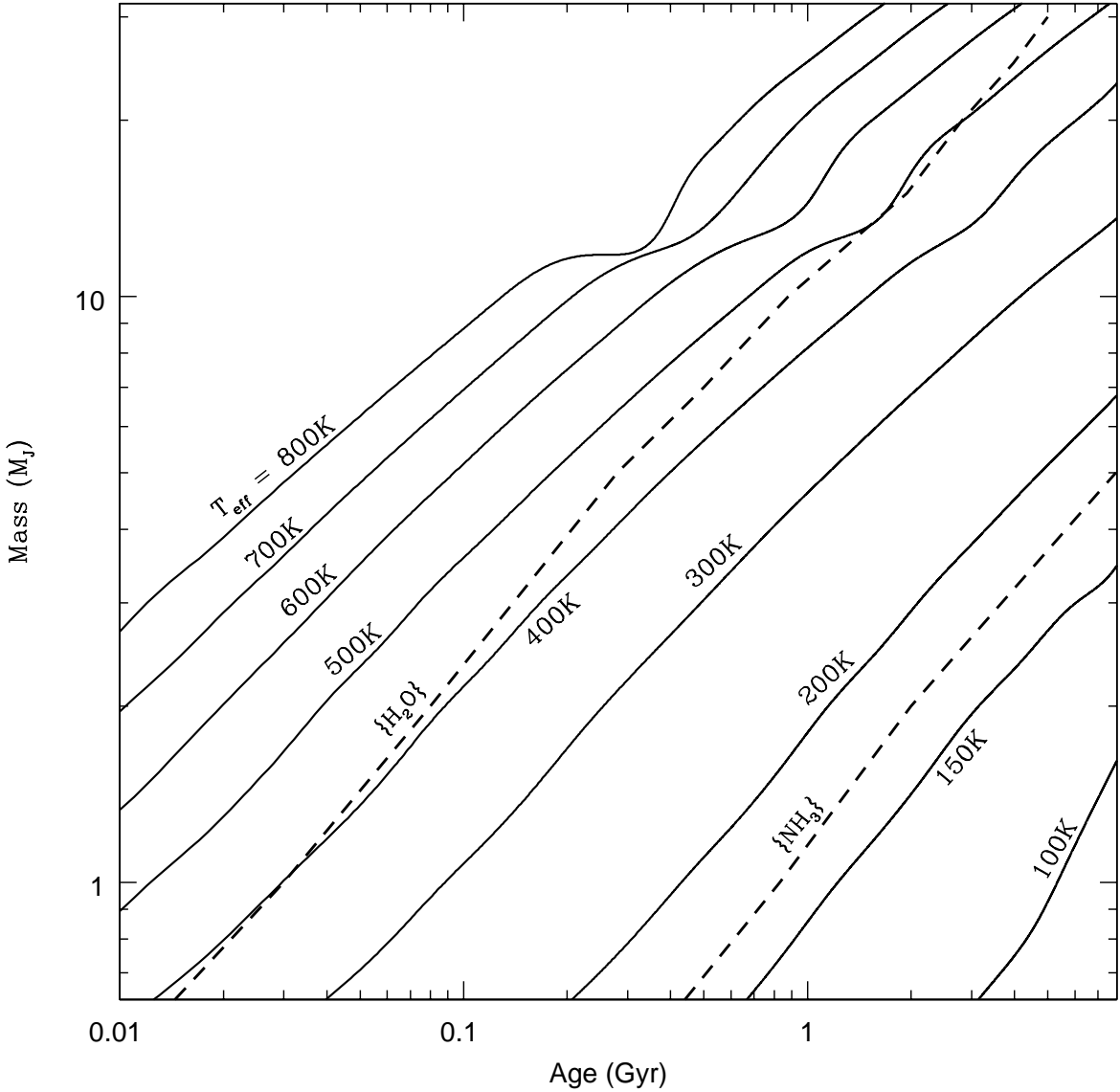}
\caption{Brown dwarf mass (in \mj) versus age (in Gyr) for a variety of
effective temperatures.  At a constant mass, incrementing
the age (i.e., reading horizontally from left to right) reveals
the decline in T$_{\textrm{eff}}$ with time.  Additionally, along such
a trajectory, the condensation curves for water and ammonia
indicate the ages at which the condensation of these species first
ensues in the outer atmosphere.  The kink near $\sim$13 M$_J$ is a
consequence of deuterium burning.}
\label{fig:2}
\end{figure}

Figure \ref{fig:1} depicts evolutionary trajectories and isochrones in \teff/$g$ space for models
in the realm beyond the T dwarfs.  The depicted isochrones span the
range from 10$^7$ to $5\times 10^9$ years and the masses cover the range from 0.5 \mj to 25 \mj.  The large dots denote 
the models found in Table 1 for which we have calculated spectra and atmospheres.  For contrast, the approximate region
in which the currently known T dwarfs reside is also shown.  In addition, we provide the demarcation lines
that separate (in a rough sense) the cloud-free models from those with water clouds and ammonia clouds.
The clouds form to the left of the corresponding condensation lines.  Figure \ref{fig:1}
emphasizes the transitional and as-yet-unstudied character of this family of objects.  It also
provides at a glance a global summary of family properties.  Figure \ref{fig:2} is a companion figure
to Fig. \ref{fig:1}, but shows iso-\teff lines in mass/age space.  For a given mass, Fig. \ref{fig:2}
allows one to determine the evolution of \teff and at what age a given \teff is achieved.
It also makes easy the determination of the combination of mass and age for which clouds form,
as well as the minimum mass for which a given \teff is reached after approximately 
the galactic disk's or the sun's age ($\sim 10^{10}$ and $\sim 4.6 \times 10^9$ years, respectively).
For instance, Fig. \ref{fig:2} shows that it takes $\sim 200$ Myr for a 2-\mj object to reach a \teff of 400 K,
that it takes the same object 1 Gyr to reach a \teff of $\sim$250 K, and that in the age of the solar system
a 2-\mj object can reach the NH$_3$ condensation line.  Similarly, 
Fig. \ref{fig:2} indicates that a 10-\mj object takes $\sim$1 Gyr
to reach a \teff of $\sim$400 K, and that it has water-ice clouds in its upper atmosphere.
Figures \ref{fig:1} and \ref{fig:2} are, therefore, useful maps of the model domain to which
the reader may want often to return.

To calculate absolute fluxes at 10 parsecs one needs the radius of the object.  We determine this
for each model in Table 1 by using a fit to the results of Burrows et al. (1997) that works
reasonably well below $\sim$25 \mj and after deuterium burning has ended:
\begin{equation}
R = 0.94 \Bigl(\frac{g}{10^5\, {\rm cm\, s^{-2}}}\Bigr)^{-0.18} \Bigl(\frac{T_{\rm eff}}{1000\, {\rm K}}\Bigr)^{0.11} R_{\rm J}\, , 
\end{equation}
where $R_{\rm J}$ is Jupiter's radius ($\sim 7.15 \times 10^9$ cm).

\section{Temperature-Pressure Profiles}
\label{profiles}

\begin{figure}
\plotone{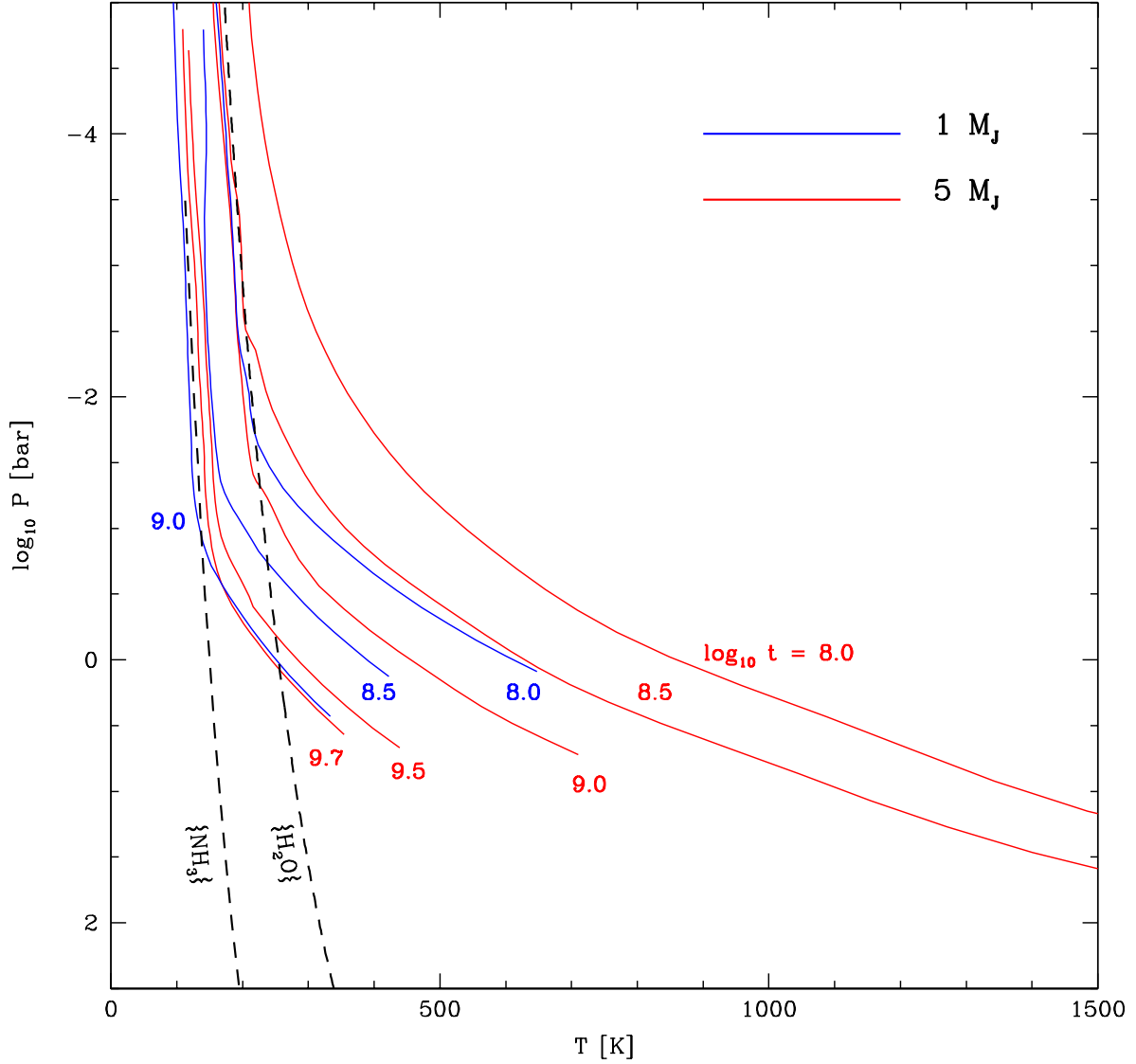}
\caption{Atmospheric temperature (T, in Kelvin) versus pressure (P, in bars) 
profiles for models with 1 (blue) and 5 (red) Jupiter masses and for
a variety of ages (in years). Also shown are the condensation curves for water and
ammonia (dashed curves).  The atmospheres of $\sim$2/3 of the objects in our model set form 
some condensed water; an isolated 1-\mj object is also expected
to contain condensed ammonia after an age of 1 billion years.}
\label{fig:4}
\end{figure}

Shown in Fig. \ref{fig:3} are representative temperature-pressure profiles at 
300 Myr (blue) for models with masses of 1, 2, 5, 7, and 10 \mj  
and at 5 Gyr (red) for models with masses of 2, 5, 7, 10, 15, 20, and 25 \mj.
Superposed are the water ice and ammonia condensation lines at solar metallicity.   
The radiative-convective boundary pressures are near 0.1-1.0 bars for the lowest-mass, oldest
models and are near 10-30 bars for the youngest, most massive models.
At a given temperature, lower-mass objects have higher pressures (at a given age).
Similarly, an object with a given mass evolves to higher and higher pressures at a given temperature.
This trend is made clear in Fig. \ref{fig:4}, in which the evolving T/P profiles for 1-\mj and 5-\mj
models are depicted, and is not unexpected (Marley et al. 1996,2002; Burrows et al. 1997). Note that 
Fig. \ref{fig:4} implies that a 5-\mj object takes $\sim$300 Myr  
to form water clouds, but that a 1-\mj object takes only $\sim$100 Myr.  After $\sim$1 Gyr, a 1-\mj object
forms ammonia clouds, signature features of Jupiter itself.  These numbers echo 
the information also found in Fig. \ref{fig:2}.

The appearance of a water-ice cloud manifests itself in Figs. \ref{fig:3} and \ref{fig:4}
by the kink in the T/P profile near the intercept with the associated condensation line.
Generally, the higher the intercept of the T/P profile with the condensation line
(the lower the intercept pressure) the smaller the droplet size (Cooper et al. 2003).   
Note that after an age of $\sim$300 Myr a 7-\mj object is expected to form water 
clouds high up in its atmosphere and that after $\sim$5 Gyr even a 25-\mj object will do so.
The higher the atmospheric pressure at which the cloud forms the greater the column
thickness of the cloud.  This results in a stronger cloud signature 
for the lower-mass models than for the higher-mass models.  
However, given the generally large ice particle
sizes derived with the Cooper et al. (2003) model, the low assumed 
supersaturation (\S\ref{approach}), the tendency for larger particle radii
to form for larger intercept pressures, and the modest to low 
imaginary part of the index of refraction for pure water ice, 
the effect of water clouds in our model set is not large.
This translates into a small cloud effect on the corresponding flux spectra (\S\ref{spectra}).

Figure \ref{fig:5} portrays the evolution of the T/P profiles for 10-\mj and 20-\mj objects.
This figure is provided to show, among other things, the position of the forsterite (Mg$_2$SiO$_4$) condensation
line relative to that of water ice.  Mg$_2$SiO$_4$ clouds exist in these brown dwarfs, 
but at significantly higher pressures and temperatures and are, therefore, buried from view.
Hence, unlike in L dwarfs, such clouds have very little effect on the emergent spectra of the coolest
brown dwarfs that are the subject of this paper.

Finally, the high pressures achieved at low temperatures for the lowest mass, oldest 
objects shown in Figs. \ref{fig:3}, \ref{fig:4}, and \ref{fig:5} suggest that
the CIA (pressure-induced) opacity of H$_2$ might for them be important.  This is indeed
the case at longer wavelengths and is discussed in \S\ref{spectra}.  We mention this because
CIA opacity is yet another characteristic signature of the Jovian planets in our own solar system
and to emphasize yet again that our cold brown dwarf model suite is a bridge between the realms of the 
planets and the ``stars."

\section{SIRTF and JWST Point-Source Sensitivities}
\label{sense}

Before we present and describe our model spectra, we discuss the anticipated
point-source sensitivities of the instruments on board the SIRTF and JWST space telescopes.  
SIRTF has a 0.84-meter aperture and is to be launched in mid-April of 2003.  JWST 
is planned to have a collecting area of $\sim$25 square meters over a segmented 6-meter diameter mirror
and is to be launched at the beginning of the next decade. 
While SIRTF is the last of the ``Great Observatories," and will view the sky with
unprecedented infrared sensitivity, JWST will in turn provide a two- to four-order-of-magnitude
gain in sensitivity through much of the mid-infrared  up to 27 microns. 
While their fields of view are limited and missions like WISE (formerly NGSS; Wright et al. 2001)
are more appropriate for large-area surveys, the extreme sensitivity of both SIRTF
and JWST will bring the coolest brown dwarfs and isolated giant planets into the realm of
detectability and study.  

SIRTF/IRAC has four channels centered at 3.63 \mic, 4.53 \mic, 5.78 \mic, and 8.0 \mic
that are thought to have 5-$\sigma$ point-source sensitivities for 200-second integrations
of $\sim$2.5, $\sim$4.5, $\sim$15.5, and $\sim$25.0 microJanskys, respectively.  HST/NICMOS
achieves a bit better than one microJansky sensitivity at 2.2 \mic, but does not extend as far into the near IR.  
The short-wavelength, low-spectral resolution module (``Short-low") of 
SIRTF/IRS extends from $\sim$5.0 \mic to $\sim$14.0 \mic and has a 5-$\sigma$ point-source
sensitivity for a 500-second integration of $\sim$100 microJanskys.   The other three modules on IRS cover 
other mid-IR wavelength regimes at either low- or high-spectral resolution, but will have 
smaller brown dwarf detection ranges.  The $\sim$20.5 \mic to $\sim$26 \mic
channel on SIRTF/MIPS is the most relevant channel on MIPS for brown dwarf studies and has a suggested 
1-$\sigma$ point-source sensitivity at $\sim$24 \mic of $\sim$70 microJanskys.  This is
$\sim$1000 times better in imaging mode than for the pioneering IRAS.   All these SIRTF sensitivities are derived from
various SIRTF web pages and are pre-launch estimates 
(\verb"http://sirtf.caltech.edu").  Furthermore, for all three SIRTF
instruments, one can estimate the point-source sensitivities 
for different values of the signal-to-noise and integration
times.  However, these signals-to-noise and integration times are the nominal combinations
for each instrument and the quoted sensitivities  
serve to guide our assessment of SIRTF's capabilities for cool brown dwarf studies in advance of
real on-orbit calibrations and measurements.  

The capabilities of JWST are even more provisional, but the design goals for its instruments
are impressive (\verb"http://ngst.gsfc.nasa.gov").  JWST/NIRCam is to span 
$\sim$0.6 \mic to $\sim$5.0 \mic in various wavelength channels/filters,
though the final design has not been frozen.  The seven so-called ``B" filters have 
widths of 0.5--1.0 microns centered at $\sim$0.71, $\sim$1.1,
$\sim$1.5, $\sim$2.0, $\sim$2.7, $\sim$3.6, and $\sim$4.4 microns and are expected to have 5-$\sigma$
point-source sensitivities in imaging mode, for an assumed exposure time of $5\times 10^4$ seconds, of $\sim$1.6,
$\sim$0.95, $\sim$1.0, $\sim$1.2, $\sim$0.95, $\sim$1.05, and $\sim$1.5 nanoJanskys (nJ), respectively.
In addition, a set of so-called ``I" filters, with about half to one quarter the spectral
width of the B filters, and sensitivites comparable to that of the B filters, are
available in the 1.5--5.0 \mic region.  Furthermore, JWST/NIRCam may have a tunable filter
to examine selected spectral regions beyond 2.5 \mic at a resolution ($R = \lambda/{\Delta\lambda}$) of $\sim$100, though at
the time of this writing the availability of such a capability remained uncertain. 
Hence, with JWST/NIRCam we enter the world of {\it nano}Jansky sensitivity.  This is greater than one 
hundred times more sensitive than HST/NICMOS at 2.2 \mic and enables one to probe deeply in space,
as well as broadly in wavelength.  JWST/MIRI spans the mid-IR wavelength range from $\sim$5.0 \mic to $\sim$27.0 \mic 
and will have in imaging mode a 10-$\sigma$ point-source sensitivity for a 10$^4$-second integration of from $\sim$63 nJ
at the shortest wavelength to $\sim$10 microJanskys at the longest.  This is orders of magnitude more sensitive
than any previous mid-IR telescope in imaging mode.  (In spectral mode with 
an $R$ near 1000, JWST/MIRI
will be $\sim$100 times less sensitive than in imaging mode.)  Given 
the importance of the mid-IR for understanding those brown
dwarfs that may exist in relative abundance at \teffs cooler than those of the currently known T dwarfs,
MIRI provides what is perhaps a transformational capability.  As with SIRTF, the quoted JWST sensitivities
are taken from the associated web pages and, hence, should be
considered tentative.

We now turn to a discussion of the spectra, spectral evolution, defining features, systematics, and diagnostics
for the cool brown dwarf models listed in Table 1 and embedded in Figs. \ref{fig:1} and \ref{fig:2}.
On each of Figs. \ref{fig:6} to \ref{fig:11} in \S\ref{spectra}, we plot for the SIRTF (red) and JWST (blue)
instruments the broadband sensitivities we have summarized in this section.

\section{Cool Brown Dwarf Spectra}
\label{spectra}

\begin{figure}
\plotone{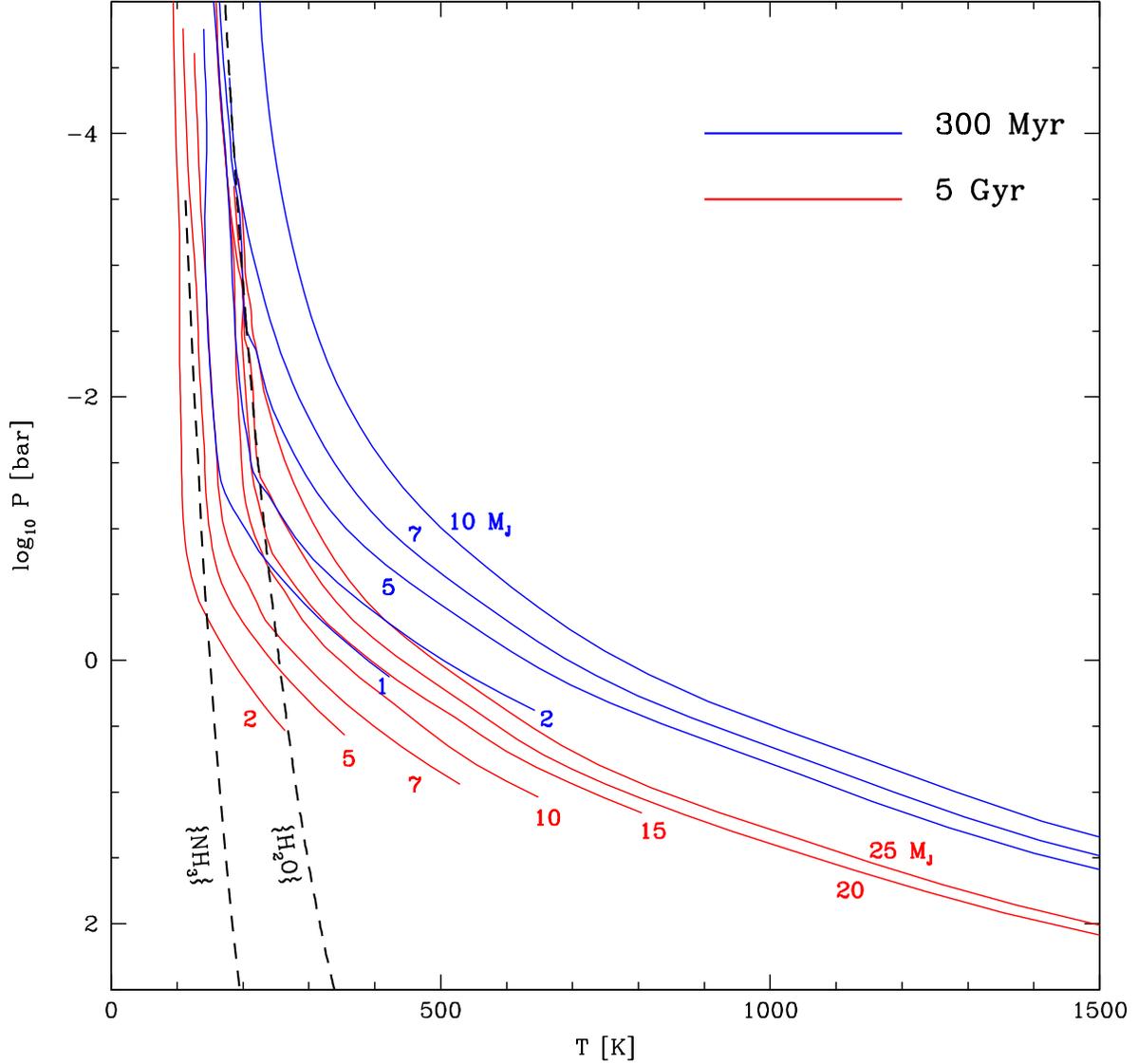}
\caption{Atmospheric temperature (T, in Kelvin) versus pressure (P, in bars) for a selection of lower-mass brown
dwarfs, both young (blue, 300 million years) and old (red, 5 billion years).  
The ordinate goes from high pressures at the bottom to low pressures at the top.
Also shown are the condensation curves for water and ammonia (dashed curves).
The presence of water clouds is expected to be important in low-mass and/or
old objects.  Ammonia clouds are likely to be relevant only in the lowest-mass,
oldest brown dwarfs.} 
\label{fig:3}
\end{figure}

\begin{figure}
\plotone{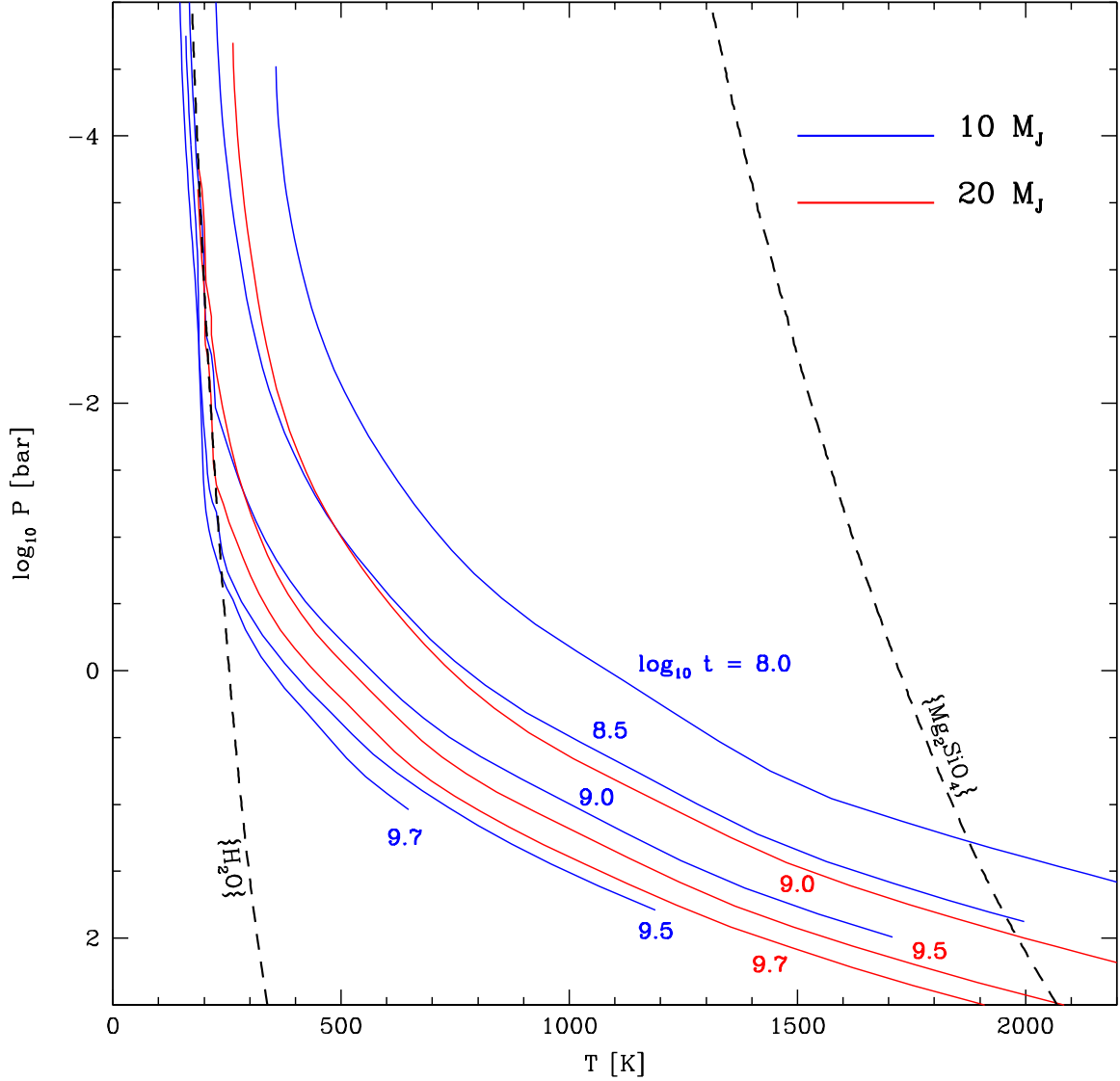}
\caption{T/P profiles, as in Figs. \ref{fig:3} and \ref{fig:4}, but for 10-\mj and 20-\mj models and for
a variety of ages (in years). Also shown are the condensation curves (dashed) for water
and forsterite, a representative silicate.  The oldest 20-\mj objects are
expected to contain some condensed water in their outer atmospheres.
Silicate layers will be buried deeply, typically at pressures near 
$\sim$100 bars.}
\label{fig:5}
\end{figure}

\begin{figure}
\plotone{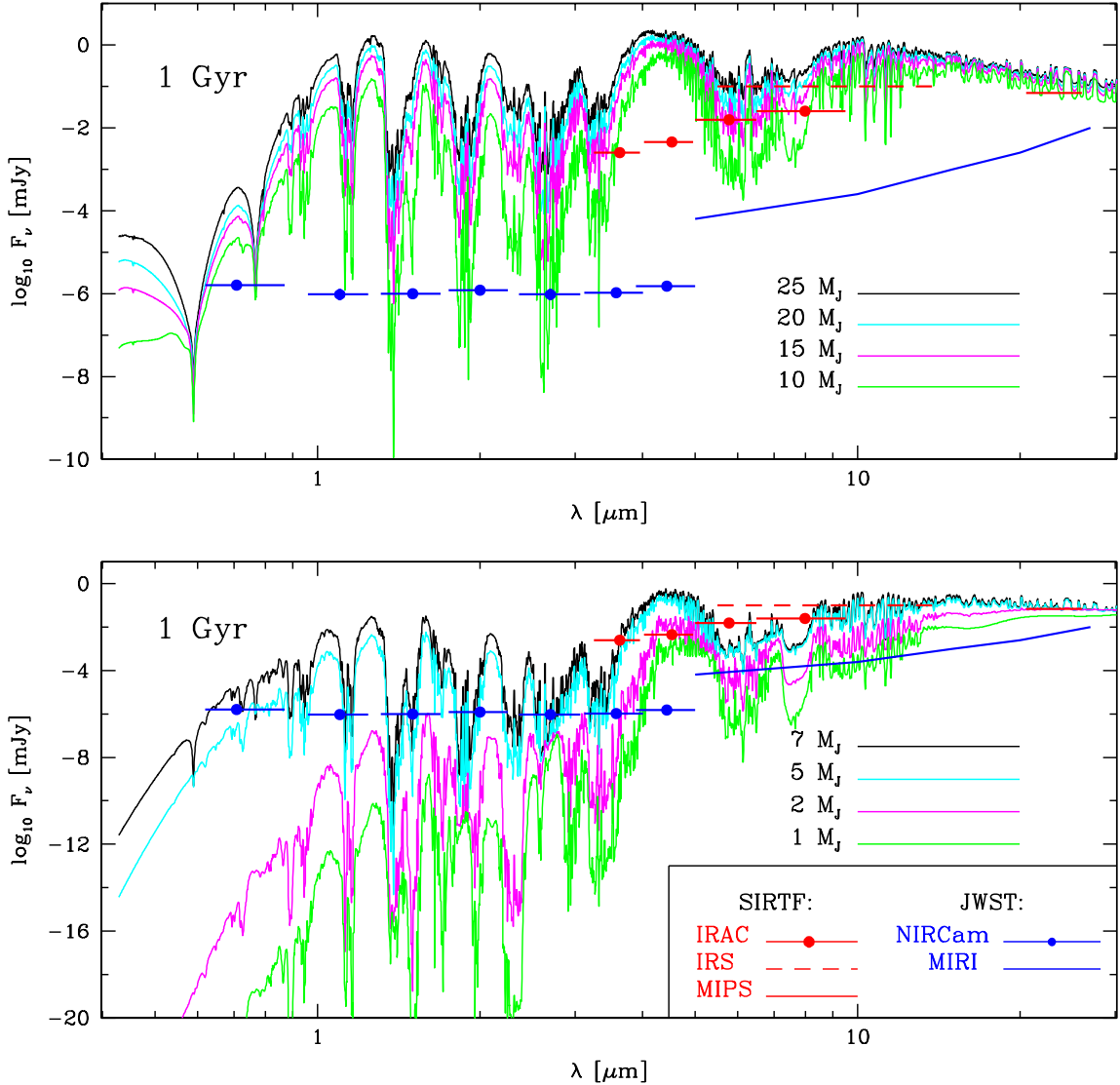}
\caption{Spectra (${F_{\nu}}$, in milliJanskys) versus wavelength (in microns)
for a range of brown dwarf masses at an age of 1
Gyr and a distance of 10 pc.  The top panel is for masses of 25, 20, 15, and 10 \mj
and the bottom panel is for masses of 7, 5, 2, and 1 \mj.  Note that the ordinate scales are different.  
Flux densities are plotted for wavelengths from $\sim$0.4 to 30 \mic.
Superposed are the approximate point-source sensitivities
for instruments on SIRTF (red) and JWST (blue) (see \S\ref{sense}).  All sensitivities are for imaging mode,
except for those of SIRTF/IRS, which are for spectral mode.  The SIRTF/IRAC sensitivities
for its channels 1 (3.63 \mic), 2 (4.53 \mic), 3 (5.78 \mic), and 4 (8.0 \mic)
are 5-$\sigma$ and are for an exposure time of 200 seconds.  The 5-$\sigma$ ``Short-low" SIRTF/IRS
sensitivities (dashed) from 5.0 to 14 \mic are for an exposure time of 500 seconds.  The SIRTF/MIPS
1-$\sigma$ sensitivities (dotted) are for wavelengths from 20.5 to 26 \mic and an exposure
time of 200 seconds.  The JWST/NIRCam sensitivities for its B1 (0.71 \mic), B2 (1.1 \mic),
B3 (1.5 \mic), B4 (2.0 \mic), B5 (2.7 \mic), B6 (3.57 \mic), and B7 (4.44 \mic)
bands are 5-$\sigma$ and assume an exposure time of $5\times 10^4$ seconds.  The JWST/MIRI
sensitivity curve from 5.0 to 27 \mic is 10-$\sigma$ and assumes an exposure time
of 10$^4$ seconds. See text for discussion.}
\label{fig:6}
\end{figure}

\begin{figure}
\plotone{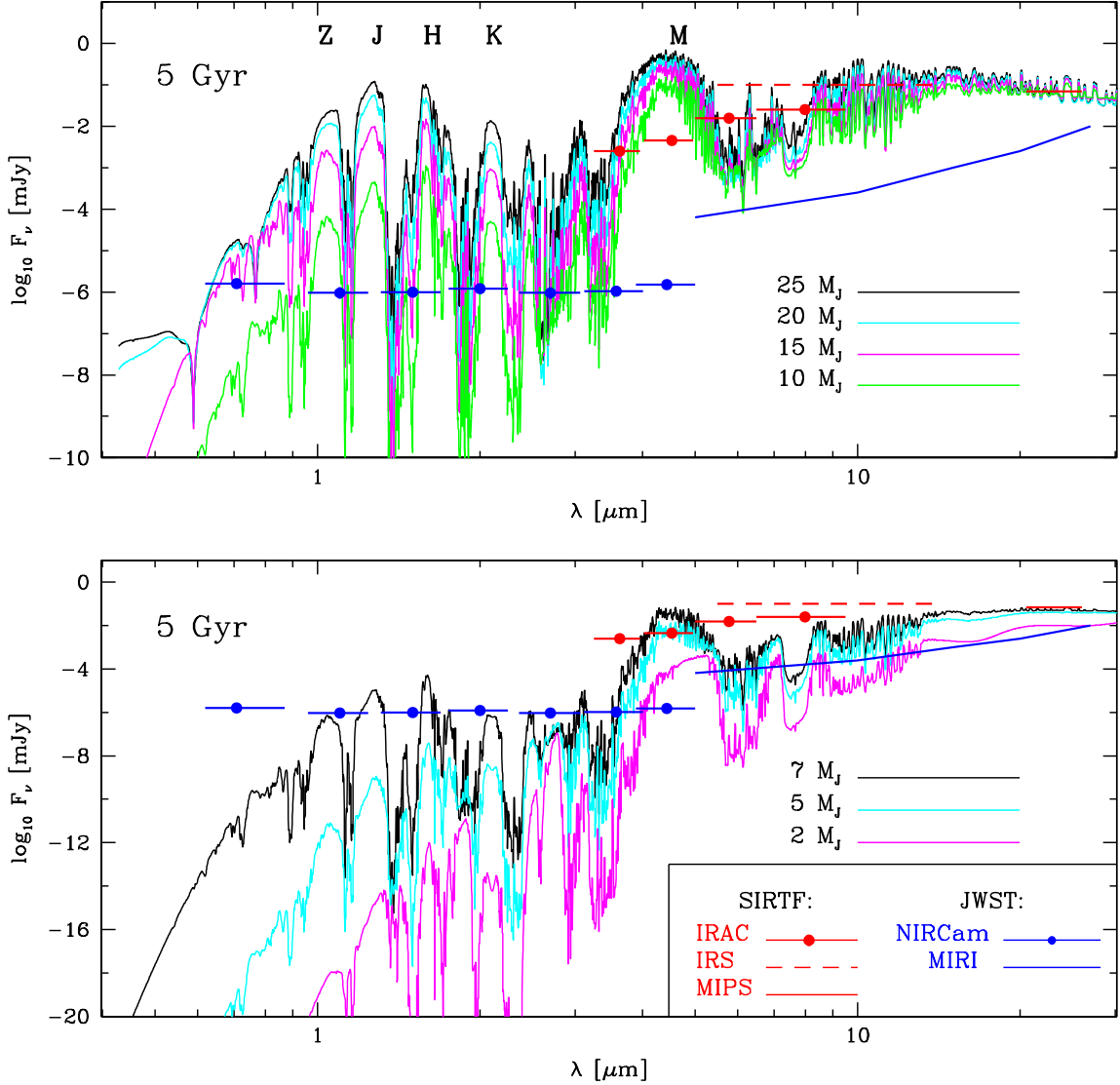}
\caption{Spectra ($F_{\nu}$, in milliJanskys) for a range of brown dwarf masses at an age of 5
billion years and a distance of 10 pc.  Flux densities 
are plotted from $\sim$0.4 to 30 \mic.  The top panel is for masses of 25, 20, 15, and 10 \mj
and the bottom panel is for masses of 7, 5, and 2 \mj.  The positions of the $Z$, $J$, $H$, $K$,
and $M$ photometric bands are indicated in the top panel.
Superposed are the suggested point-source sensitivities
for instruments on SIRTF (red) and JWST (blue) that are described in \S\ref{sense}
and in the caption to Fig. \ref{fig:6}.}
\label{fig:7}
\end{figure}

\begin{figure}
\plotone{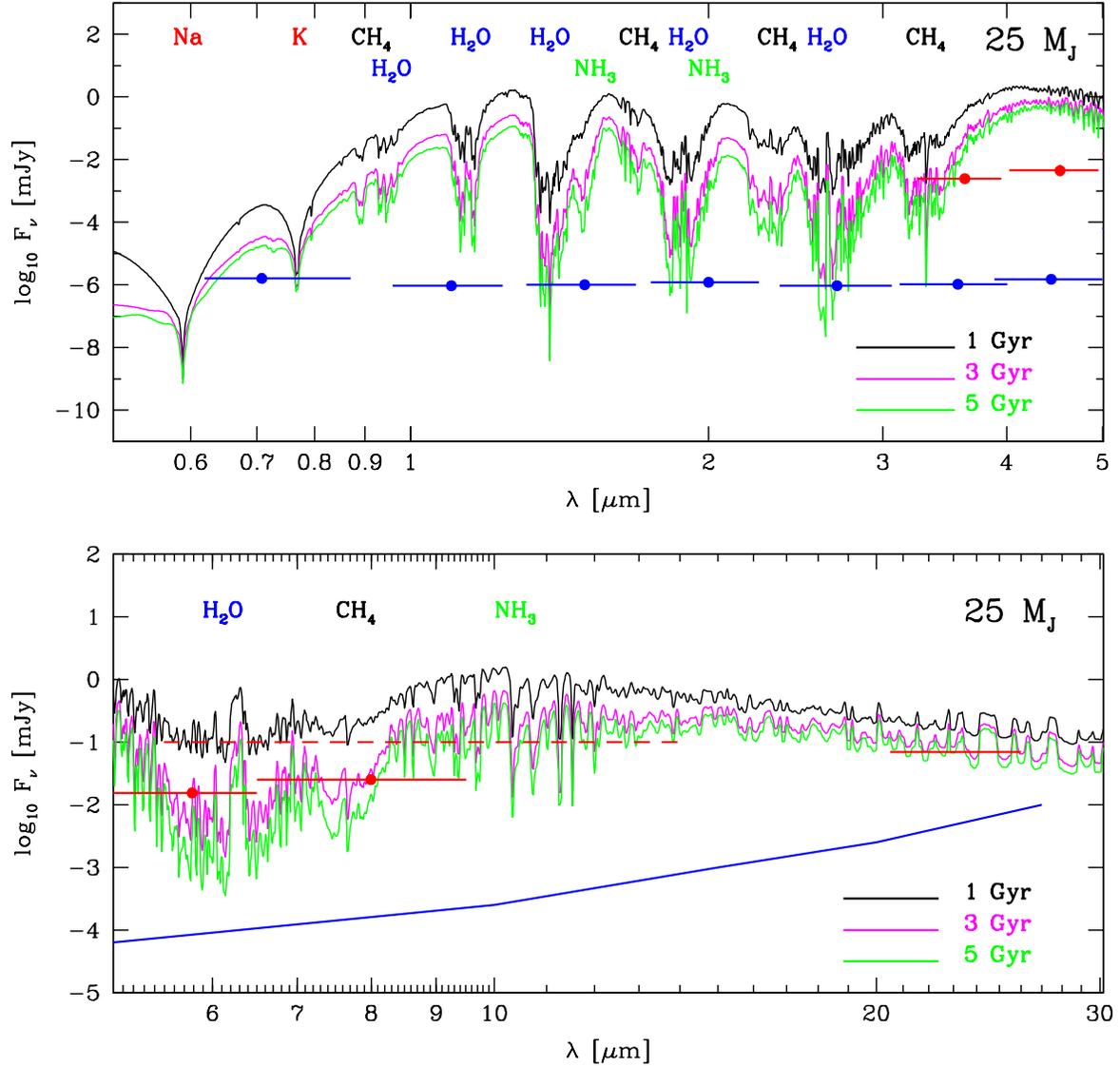}
\caption{Spectra ($F_{\nu}$ in milliJanskys versus wavelength in microns) of 25-\mj brown dwarfs at 1, 3,
and 5 billion years.  As in Figure \ref{fig:6}, SIRTF and
JWST detector sensitivities are also plotted (see Figure
\ref{fig:6} for legend).  The top panel is from 0.5 \mic to 5 \mic and the bottom 
panel is from 5 \mic to 30 \mic.  Gaseous water absorption
features and the pressure-broadened sodium and potassium resonance
lines remain strong over this range of ages, while methane and ammonia features
strengthen with age. The wavelength positions of various of the molecular 
and atomic features are depicted for reference on this and subsequent
figures. See text for details.}  
\label{fig:8}
\end{figure}

\begin{figure}
\plotone{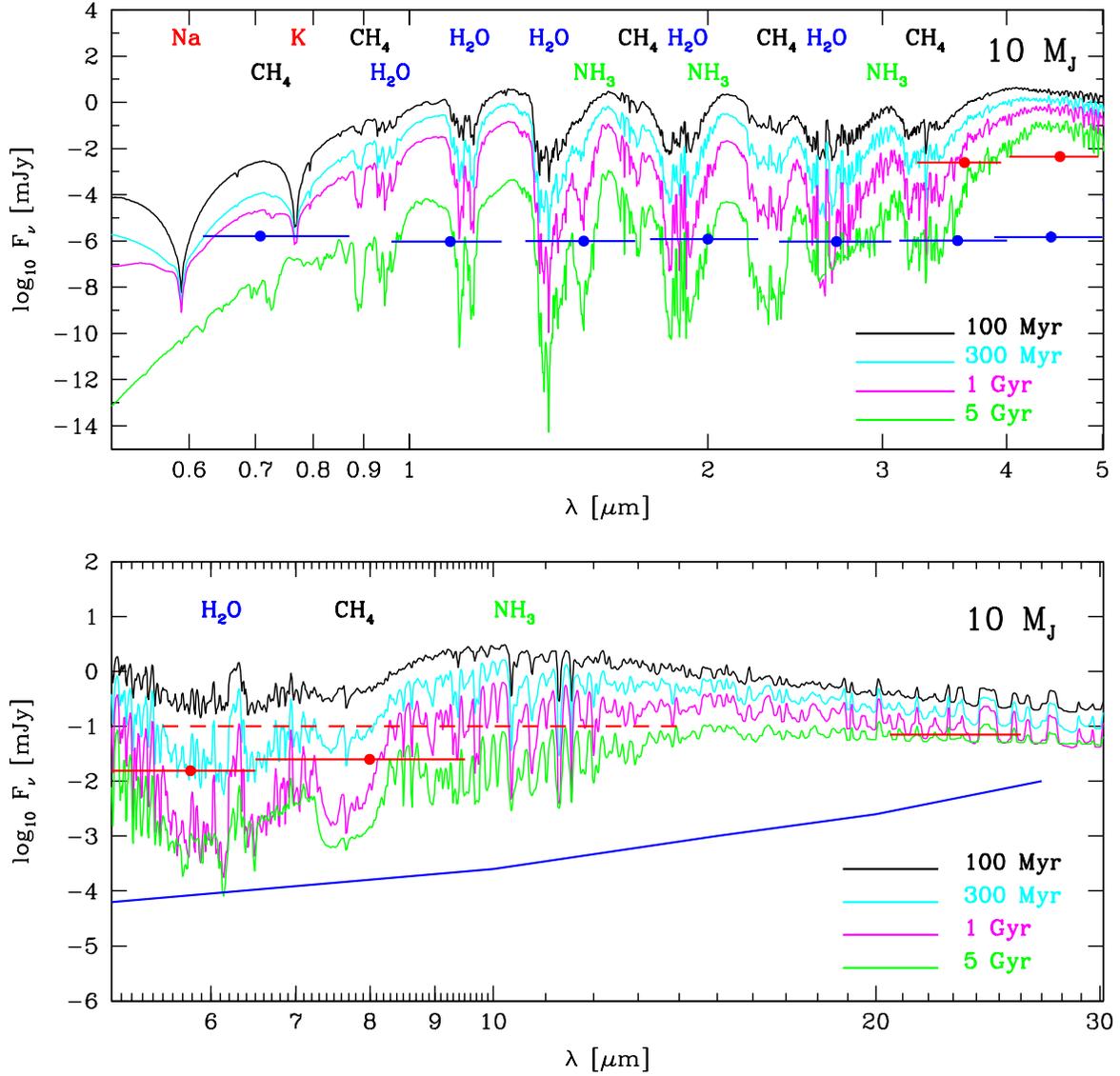}
\caption{Similar to Fig. \ref{fig:8}, but for 10-\mj brown dwarfs at $10^8, 3 \times  10^8,
10^9,$ and $5 \times  10^9$ years.  As in Figure \ref{fig:6},
SIRTF and JWST detector sensitivities are plotted (see Figure
\ref{fig:6} for legend).  The top panel is from 0.5 \mic to 5 \mic and the bottom
panel is from 5 \mic to 30 \mic.  Methane and ammonia absorption
features strengthen with age, as the alkali lines wane.}
\label{fig:9}
\end{figure}

\begin{figure}
\plotone{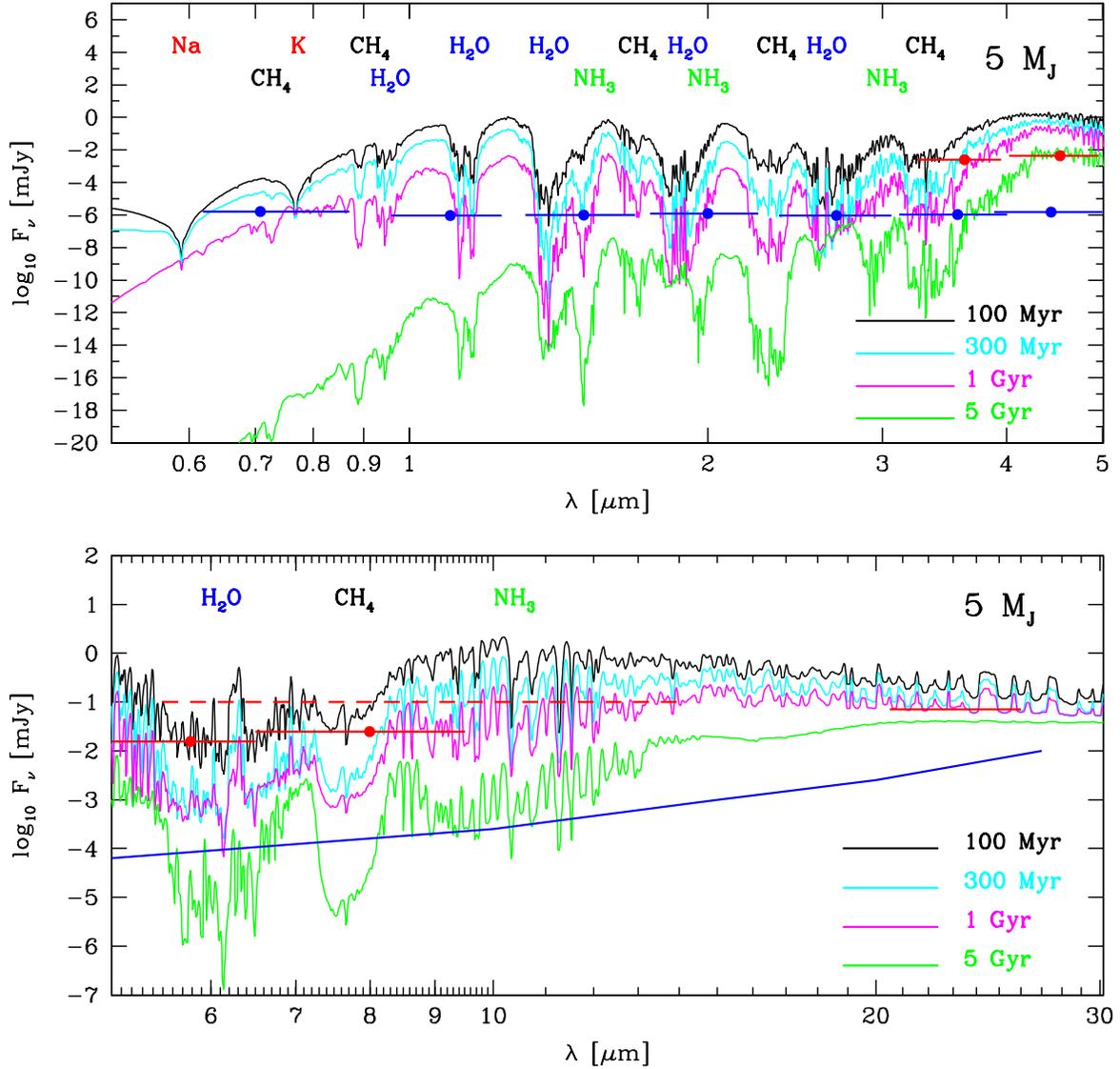}
\caption{Similar to Fig. \ref{fig:8}, but for 5-\mj brown dwarfs at $10^8, 3 \times  10^8,
10^9,$ and $5 \times  10^9$ years.  Flux densities are in milliJanskys and the 
wavelengths are from 0.5 \mic to 5 \mic in the top panel 
and from 5 \mic to 30 \mic in the bottom panel.  As in Figure \ref{fig:6},
SIRTF and JWST detector sensitivities are plotted (see Figure
\ref{fig:6} for legend).  Methane and ammonia absorption
features strengthen with age.}
\label{fig:10}
\end{figure}

\begin{figure}
\plotone{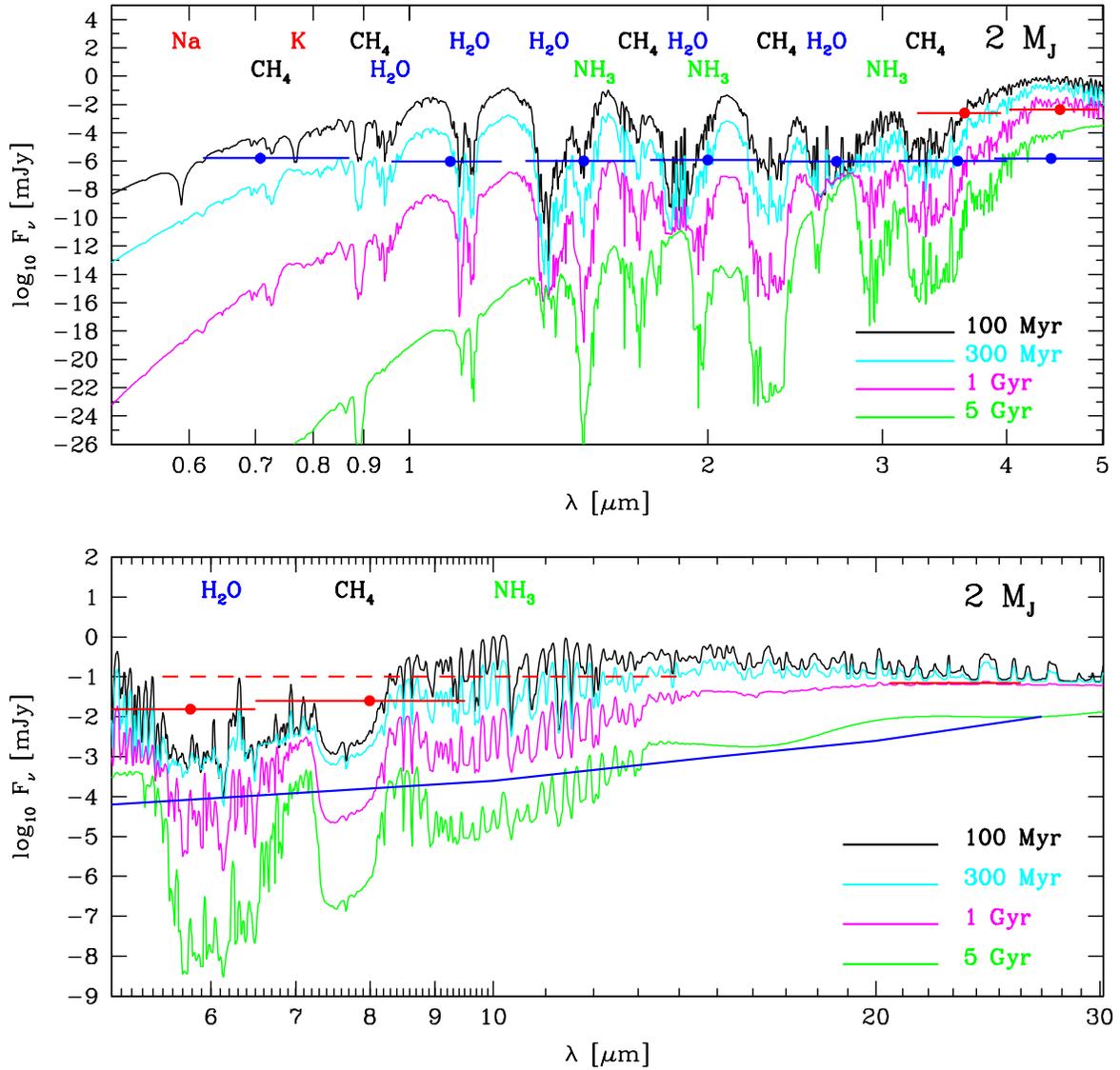}
\caption{Similar to Figs. \ref{fig:8} through \ref{fig:10}, but for 2-\mj brown dwarfs at $10^8, 3 \times 10^8,
10^9,$ and $5 \times  10^9$ years.  As in Figure \ref{fig:6},
SIRTF and JWST detector sensitivities are plotted (see Figure
\ref{fig:6} for legend).  The top panel is from 0.5 \mic to 5 \mic and the bottom
panel is from 5 \mic to 30 \mic.  Methane absorption features
are strong over this full range of ages, while the ammonia features strengthen
with age.  The sodium and potassium resonance lines
are prominent in only the youngest objects.}
\label{fig:11}
\end{figure}

Using the numerical tools and data referred to in \S\ref{approach}, and the mapping between
\teff/$g$ and mass/age found in Table 1, we have generated a grid of spectral and atmospheric 
models for cool brown dwarfs that reside in the low-\teff sector of \teff/$g$ space (Fig. \ref{fig:1}).
Some of the associated T/P profiles were given in Figs. \ref{fig:3}, \ref{fig:4}, and \ref{fig:5}.
In  Figs. \ref{fig:6} to \ref{fig:11}, we plot theoretical flux spectra ({F}$_{\nu}$, in milliJanskys) 
from the optical to 30 \mic at a distance of 10 parsecs.  These figures constitute the major results 
of our paper. For comparision, superposed on each figure are the estimated point-source
sensitivities of the instruments on board SIRTF and JWST (\S\ref{sense}).  In addition, included at the top of Figs. \ref{fig:8}
through \ref{fig:11} are the rough positions of the major atmospheric absorption features. 
(The full model set is available from the first author upon request.)

Figures \ref{fig:6} and \ref{fig:7} portray the mass dependence of a cool brown dwarf's flux spectrum 
at 10 parsecs for ages of one and five Gyr, respectively.  The model masses are 25, 20, 15, 10, 7, 5, 2, and 1 \mj.
The top panels depict the most massive four, while the bottom panels depict the least massive four (three for Fig. \ref{fig:7}).
Together they show the monotonic diminution of flux with object mass at a given age that parallels the
associated decrease in \teff with mass (from $\sim$800 K to $\sim$130 K) seen in Table 1 and Fig. \ref{fig:2}.

Figures \ref{fig:6} through \ref{fig:11} show the peaks due to enhanced flux through the water vapor
absorption bands that define the classical terrestrial photometric bands ($Z$, $J$, $H$, $K$, and $M$) and 
that have come to characterize brown dwarfs since the discovery of Gliese 229B (Oppenheimer et al. 1995; Marley et al. 1996).
For the more massive models, the near-IR fluxes are significantly 
above black-body values.  At \teff{}$\sim$800 K, the 25-\mj/1-Gyr
model shown in Fig. \ref{fig:6} could represent the known late T dwarfs, but all other models 
in this model set are ``later" and, hence, represent as yet undetected objects.   

Apart from the distinctive water troughs, generic features are the hump at 4-5 microns ($M$ band), the broad
hump near 10 microns, the methane features at 2.2 \mic, 3.3 \mic, 7.8 \mic, and in the optical (particularly
at 0.89 \mic), the ammonia features at $\sim$1.5 \mic, $\sim$1.95 \mic, $\sim$2.95 \mic, and $\sim$10.5 \mic,
and the Na-D and K I resonance lines at 0.589 \mic and 0.77 \mic, respectively.  However, as Figs. \ref{fig:6}-\ref{fig:11}
indicate, the strengths of each of these features are functions of mass and age.

For lower masses or greater ages, the centroid of the $M$ band hump shifts from $\sim$4.0 \mic to
$\sim$5.0 \mic.  In part, this is due to the swift decrease with \teff at the shorter wavelengths of the Wien tail.
Even after the collapse of the flux in the optical and near-IR after $\sim$1 Gyr for masses below
5 \mj or after $\sim$5 Gyr for masses below 10 \mj, the $M$ band flux persists as a characteristic
marker and will be SIRTF's best target.  Moreover, IRAC's filters are well-positioned for this task. 
As one would expect, the relative importance of the 
mid-IR fluxes, in particular between 10 and 30 microns, grows with decreasing mass and increasing age.
Since this spectral region is near the linear Rayleigh-Jeans tail, fluxes here persist despite decreases
in \teff from $\sim$800 K to $\sim$130 K. Figure \ref{fig:11} depicts this clearly for the older 2-\mj models.
The rough periodicity in flux beyond 10 \mic is due predominantly to the presence of pure rotational bands of water and, for
cooler models, methane as well. For the coldest models depicted in Figs. \ref{fig:6}, \ref{fig:7}, and \ref{fig:11},
this behavior subsides, but is replaced with long-period undulations due to CIA absorption by H$_2$.
Such a signature is characteristic of Jovian planets and is expected for low-T, high-P
atmospheres.  Its appearance marks yet another transition, seen first in this model set for the old 5-\mj and middle-aged
2-\mj objects, between T-dwarf-like and ``planet"-like behavior.  As Figs. 
\ref{fig:6}-\ref{fig:11} imply, SIRTF/MIPS should be able to
detect at 10 parsecs the $\sim$24-\mic flux of objects more massive than 2-4 \mj at age 1 Gyr  
or more massive than 10 \mj at 5 Gyr.

Methane forms at low temperatures and high pressures and makes its presence felt
in older and less massive objects.  Hence, its features at 0.89 \mic, 2.2 \mic, 3.3 \mic,
and 7.8 \mic deepen with age and decreasing mass. An example of such strengthening at 7.8 \mic and 2.2 \mic can be seen
in Fig. \ref{fig:10} by comparing the 100-Myr and 5-Gyr models with a mass of 5 \mj.
Clear indications of the strengthening of the methane absorption feature at 0.89 \mic with decreasing mass can be seen 
in the upper panel of Fig. \ref{fig:7}.   This trend is accompanied by a corresponding weakening of the 
Cs I feature on top of it.  However, due to its presence in the $I$ band at relatively short
wavelengths, the methane feature at 0.89 \mic may be difficult to detect for all but the youngest and/or
most massive models.  The actual strength of the 7.8-\mic feature depends on the T/P profile
in the upper layers of the atmosphere, which in turn might be 
affected by ambient UV (disfavored for free-floating
brown dwarfs) or processes
that could create a stratosphere and a temperature inversion.  
Hence, the filling in or reshaping of the 7.8-\mic feature 
might signal the presence of a stratosphere.  Such a temperature inversion could also affect 
the depths of the water troughs.

As can be seen by comparing the top panels of Figs. \ref{fig:8}-\ref{fig:11}, the  
alkali metal features at 0.589 \mic and 0.77 \mic diminish in strength with decreasing
mass and increasing age.  These features are signatures of the known T dwarfs (Burrows, Marley, and Sharp 2000;
Burrows et al. 2002; Tsuji, Ohnaka, and Aoki 1999), so their decay signals a gradual transformation
away from standard T-dwarf behavior.  For the 10-\mj model older than 1 Gyr and the 
2-\mj model older than 100 Myr, these alkali resonance features cease to be primary signatures. 
This happens near a \teff of 450 K.

Ammonia makes an appearance at even lower temperatures than methane and due to the relatively
high abundance of nitrogen its absorption features are generally strong, particularly for the cool objects in our model set.
For the higher \teffs in the mid-T-dwarf range, ammonia may have been seen, but is weak (Saumon et al. 2000). 
Figs. \ref{fig:10} and \ref{fig:11} evince strong ammonia features 
in the upper panels at $\sim$1.5 \mic, $\sim$1.95 \mic, and $\sim$2.95 \mic and in Figs. \ref{fig:8}-\ref{fig:11}
in the lower panels at $\sim$10.5 \mic.  As Figs. \ref{fig:6}-\ref{fig:11}
imply, the Short-low module on SIRTF/IRS should be able to study the 10.5-\mic ammonia feature.  
Even for the 25-\mj/1Gyr model, the  $\sim$10.5 \mic feature is prominent.
For the more massive objects (10-25 \mj), the strength of the 10.5-\mic feature increases with age.
For the lowest mass objects (2-7 \mj), the strength of the 10.5-\mic ammonia feature actually 
decreases with age, even though the strengths of the other ammonia lines increase.  
As the more massive objects age, their atmospheric
pressures increase, shifting the N$_2$/NH$_3$ equilibrium towards NH$_3$.  For the less
massive models, pressured-induced absorption by H$_2$ grows with increasing atmospheric
pressure (Fig. \ref{fig:3}-\ref{fig:4}) and partially flattens an otherwise strengthening 
10.5-\mic ammonia feature.

Below \teffs of $\sim$160 K, Figs. \ref{fig:1} and \ref{fig:2} demonstrate that ammonia clouds form.
However, given that we are studying isolated objects that have no reflected 
component (unlike Jupiter and Saturn), and given that realistic supersaturations are only $\sim$1\%, we have determined  
that ammonia clouds do not appreciably affect the emergent spectra.  As a consequence, we ignore 
them in the three relevant models (Fig. \ref{fig:1}).

As with the known T and L dwarfs, water vapor absorptions dominate and sculpt the flux spectra of the cooler brown dwarfs
and these features generally deepen with increasing age and decreasing mass.  The latter trend is in part a consequence
of the increase with decreasing gravity of the column depth of water above the (roughly-defined) photosphere.
At \teffs below $\sim$400-500 K (Figs. \ref{fig:1}-\ref{fig:5}), water condenses in brown dwarf atmospheres.  
The appearance of such water-ice clouds constitutes yet another milestone along the bridge from
the known T dwarfs to the giant planets.  Associated with cloud formation is the depletion of water vapor
above the tops of the water cloud, with the concommitant decrease at altitude in the gas-phase abundance 
of water.  Within $\sim$100 Myr, water clouds form in the atmosphere of
an isolated 1-\mj object and within $\sim$5 Gyr they form in the atmosphere of a 25-\mj object.  In fact, 
approximately two-thirds of the models listed in Table 1 incorporate water-ice clouds.  However, at supersaturations
of 1\% and for particle sizes above 10 microns (\S\ref{approach}-\S\ref{profiles}; 
Cooper et al. 2003), such clouds (and the corresponding
water vapor depletions above them) only marginally
affect the calculated emergent spectra.  Even though we see in Figs. \ref{fig:3}-\ref{fig:5}
the associated kinks in the T/P profiles, these do not translate into a qualitative change in the 
emergent spectra at any wavelength.  For wavelengths longward of 1 micron, the cloudy spectra differ from the no-cloud
spectra by at most a few tens of percent.   For a representative 2-\mj model at 300 Myr (\teff{}$\sim$280 K), if we 
increase the supersturation factor by a factor of ten from 1\% to 10\%,
the flux at 5 microns decreases by approximately a factor of two, while the flux from 10 to 30 microns
increases by on average $\sim$50\%.  These are not large changes, given the many orders of magnitude
covered by the fluxes in Figs. \ref{fig:6}-\ref{fig:11}.

\begin{figure}
\plotone{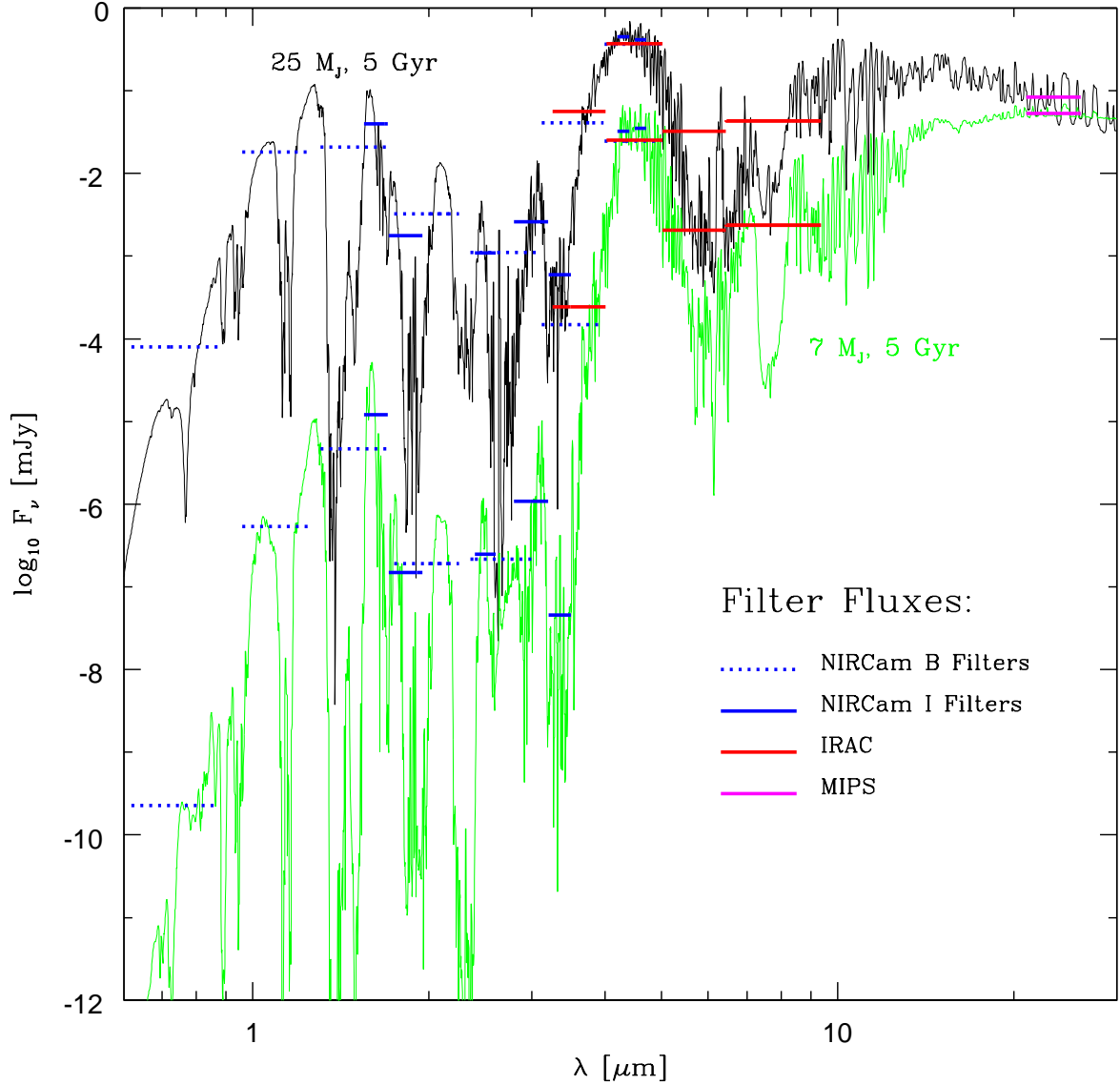}
\caption{Broadband fluxes (in milliJanskys) versus wavelength 
(in microns) at 10 parsecs for 25-\mj and 7-\mj
models at 5 Gyr predicted for SIRTF/IRAC, SIRTF/MIPS, and the NIRCam B and I filters.    
Also plotted are the corresponding model spectra (see Fig. \ref{fig:7}) from
0.6 to 30 \mic.}      
\label{fig:12}
\end{figure}

The prominence of water features provides a guide to the optimal placement of
NIRCam filters for the detection and characterization of brown
dwarfs. For example, the water feature near 0.93 \mic
is missed by the B filters, while those features at $\sim$1.4 and $\sim$1.8
\mic are not centered on the respective adjacent
filters and, hence, are diluted by the adjoining continuum. The I filters on NIRCam would partially
overcome these limitations. Even so, as Fig. \ref{fig:12} shows,
the broadband fluxes in the NIRCam filters provide useful diagnostics of
the differences among brown dwarfs and extrasolar giant planets (here
expressed as mass at a given age), with particular sensitivity to the large
flux differences between the 5-\mic window and the region shortward. A
tunable filter could provide even greater diagnostic capability by
permitting in and around the 5-\mic window a spectral resolution near 100 to
more definitively characterize the effective temperature and, hence, the mass 
of detected objects (for a given age and composition).
Nevertheless, Fig. \ref{fig:7} indicates that at 10 parsecs even a 7-\mj object at 5 Gyr should easily be detected
in imaging mode in the $J$ and $H$ bands.  In the $M$ band, a 2-\mj object could be seen by NIRCam out to $\sim$100 parsecs. 
Furthermore, a 25-\mj object at 5 Gyr and a distance of 1000 parsecs 
should be detectable by NIRCam in a number of its current broadband filters.

\begin{figure}
\plotone{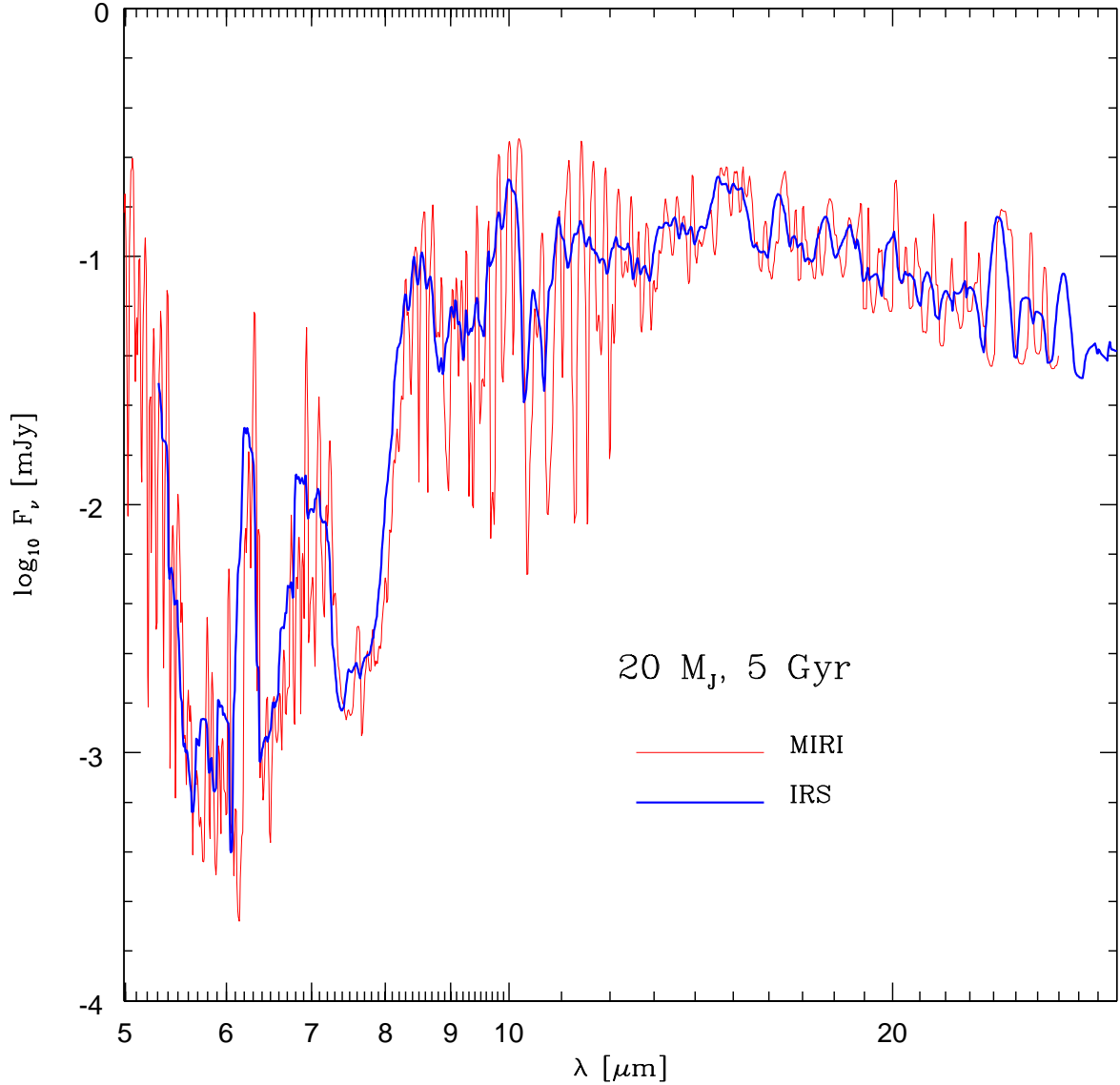}
\caption{The flux spectra (in milliJanskys) versus wavelength (in microns) 
at 10 parsecs of a 20-\mj, 5-Gyr model deresolved to
the SIRTF/IRS (blue) and JWST/MIRI (red) resolutions of 100 and 1000, respectively.
With either instrument, the water feature at $\sim$6.5 \mic, the methane feature at $\sim$7.8 \mic,
and the ammonia feature at $\sim$10.5 \mic are easily discernible, given
adequate signal-to-noise. JWST/MIRI, with its higher spectral resolution,
provides more detailed information on band positions and shapes, and, hence,
on surface gravity, temperature profile, and composition.}
\label{fig:13}
\end{figure}

Figure \ref{fig:13} shows predicted spectra of a 20-\mj/5-Gyr model in the mid-infrared
for the SIRTF/IRS and JWST/MIRI instruments.
To generate the SIRTF/IRS curve in Fig. \ref{fig:13}, we multiplied the theoretical spectra by the
IRS response curves for the entire wavelength range, not just the 5-14 \mic of the
``Short-low" module.  The IRS spectral resolution has been assumed to be 100, while that
of JWST/MIRI is $\sim$1000.  We find that the IRS spectra are useful at 10
parsecs only for the warmer brown dwarfs ($> 15$ \mj), but for these brown
dwarfs even at this modest spectral resolution one can clearly identify 
the various dominant molecular bands.

In its broadband detection (imaging) mode, JWST/MIRI will be $\sim$100 times more capable 
than SIRTF from $\sim$5 \mic to $\sim$27 \mic (\S\ref{sense}).  Since 
the mid-IR is one of the spectral regions of choice for the study of the coolest brown dwarfs, 
MIRI will assume for their characterization a role of dramatic importance.  
At wavelengths longward of 15-\mic, MIRI will be able to detect objects 10 parsecs away down to 2 \mj or lower.
In addition, it could detect an object just 10 times the
mass of Jupiter with an age of 5 Gyr out to a distance of one kiloparsec.
Furthermore, JWST/MIRI provides 10 times better spectral resolution 
than SIRTF/IRS for objects down to 10 \mj.

\begin{figure}
\plotone{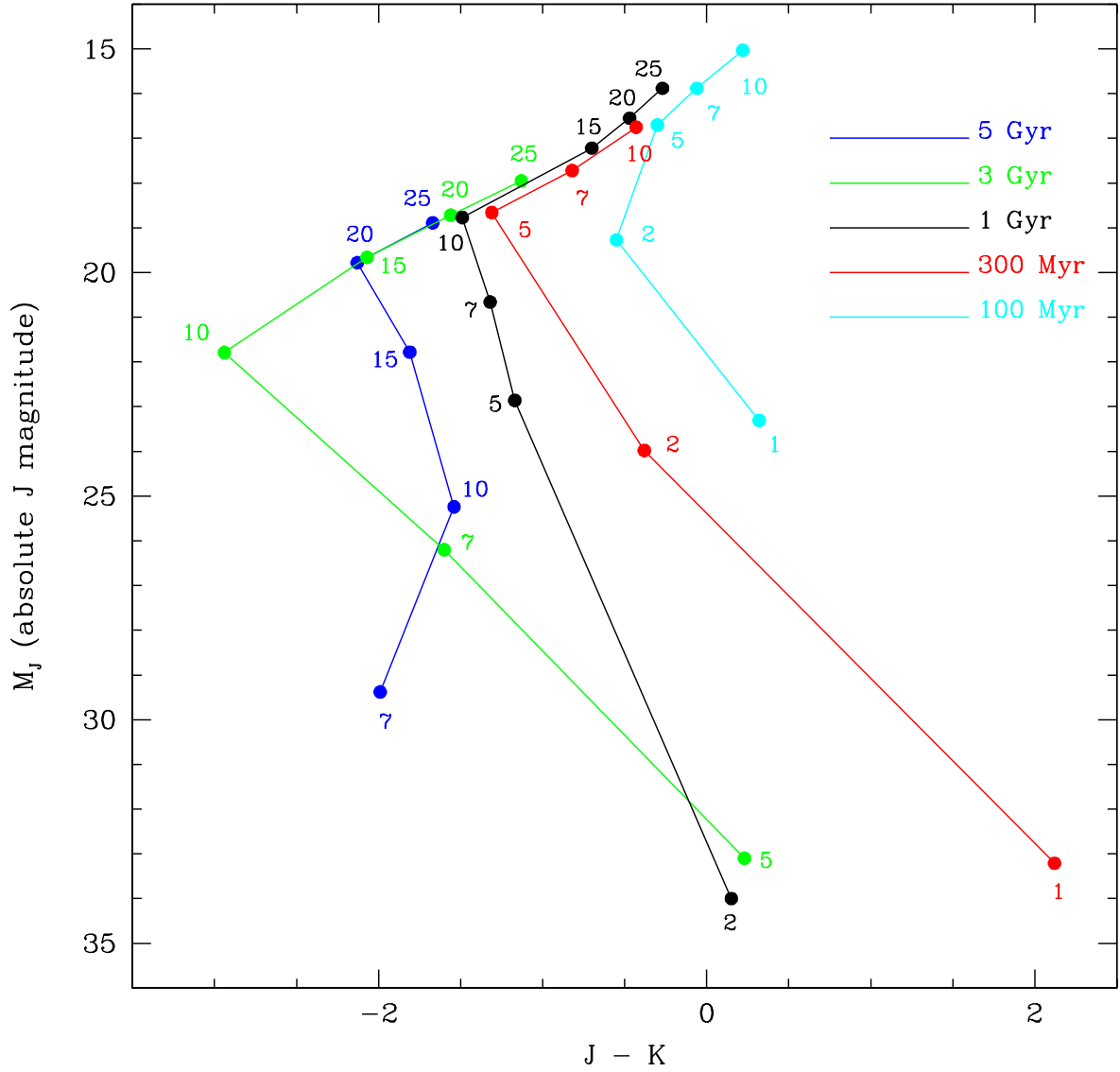}
\caption{Absolute $J$ magnitude (M$_J$) versus $J-K$ color for a range of masses and
ages. The Bessell color system and filter functions have been used (Bessell and Brett 1988).  
The numbers by the symbols denote the masses of the objects in Jupiter mass units (\mj).}
\label{fig:14} 
\end{figure}

Figs. \ref{fig:6} through \ref{fig:13} 
collectively summarize the flux spectra and evolution of the cool brown dwarfs 
yet to be discovered, as well as the extraordinary capabilities
of the various instruments on board both SIRTF and JWST for the diagnosis and characterization of their 
atmospheres.  These figures highlight the prominent molecular features of H$_2$O, CH$_4$, NH$_3$, in particular,
that are pivotal in the evolution of the differences between the coolest brown dwarfs and 
the known T dwarfs, most of which are at higher \teffs and gravities.  The latest known T dwarf
has been typed a T8 (Burgasser et al. 2000a), but its effective temperature is near 750-800 K (Geballe et al. 2001; 
Burrows et al. 2002).  This does not leave much room for the expansion of the T dwarf subtypes to the lower
\teffs and masses discussed in this paper, and suggests that yet another spectroscopic class
beyond the T dwarfs might be called for.  Many of the spectral trends described in this paper
are gradual, but the near disappearance of the alkali features below \teff = 500 K, the onset of water
cloud formation below \teff = 400-500 K, the collapse below $\sim$350 K 
of the optical and near-IR fluxes relative to those 
longward of $\sim$5 \mic, and the growing strengths of the NH$_3$ features all suggest physical
reasons for such a new class. Figure \ref{fig:14} depicts isochrones 
from 100 Myr to 5 Gyr on the $M_J$ versus $J-K$ color-magnitude
diagram and demonstrates that the blueward trend in $J-K$ that 
so typifies the T dwarfs stops and turns around (Marley et al. 2002; Stephens, Marley, and Noll 2001)
between effective temperatures of 300 and 400 K. This is predominantly due not to the appearance of water clouds,
but to the long-expected collapse of flux on the Wien tail.  
Note that the \teff at which the $J-K$ color turns around is not
the same for all the isochrones.  This is because the colors are not functions of just
\teff, but of gravity as well.  
The decrease in \teff, that for the T dwarfs squeezes
the $K$-band flux more than the $J$-band flux, finally does to $J-K$ 
what people had expected such a decrease to do 
before the discovery of T dwarfs, i.e., redden the color.  
We remind the reader that unlike M dwarfs, the $J-K$ colors
of T dwarfs actually get bluer with decreasing \teff (for a given surface gravity).  
This may be counterintuitive, but it is a result of the increasing
role of methane and H$_2$ collision-induced absorption with
decreasing temperature, as well as the positive slope of
the opacity/wavelength curve of water and its gradual steepening
with decreasing temperature. 

Were it not for the extremely 
low fluxes at such low \teffs shortward of 4 microns, we might have suggested the use of this turnaround 
to mark the beginning of a new spectroscopic class.   Moreover, clearly the optical can not be used and with
the diminishing utility of the near infrared as \teff drops, that leaves the mid-IR longward of $\sim$4 \mic
as the most logical part of the spectrum with which to characterize a new spectroscopic class. 
As is usual, this will be determined observationally, and it might be done arbitrarily to limit
the growth of the T sequence.  Nevertheless, we observe that the \teff region between 300 K and 500 K
witnesses a few physical transitions that might provide a natural break between ``stellar" types.

\section{Conclusions}
\label{conclusion}

\begin{figure}
\plotone{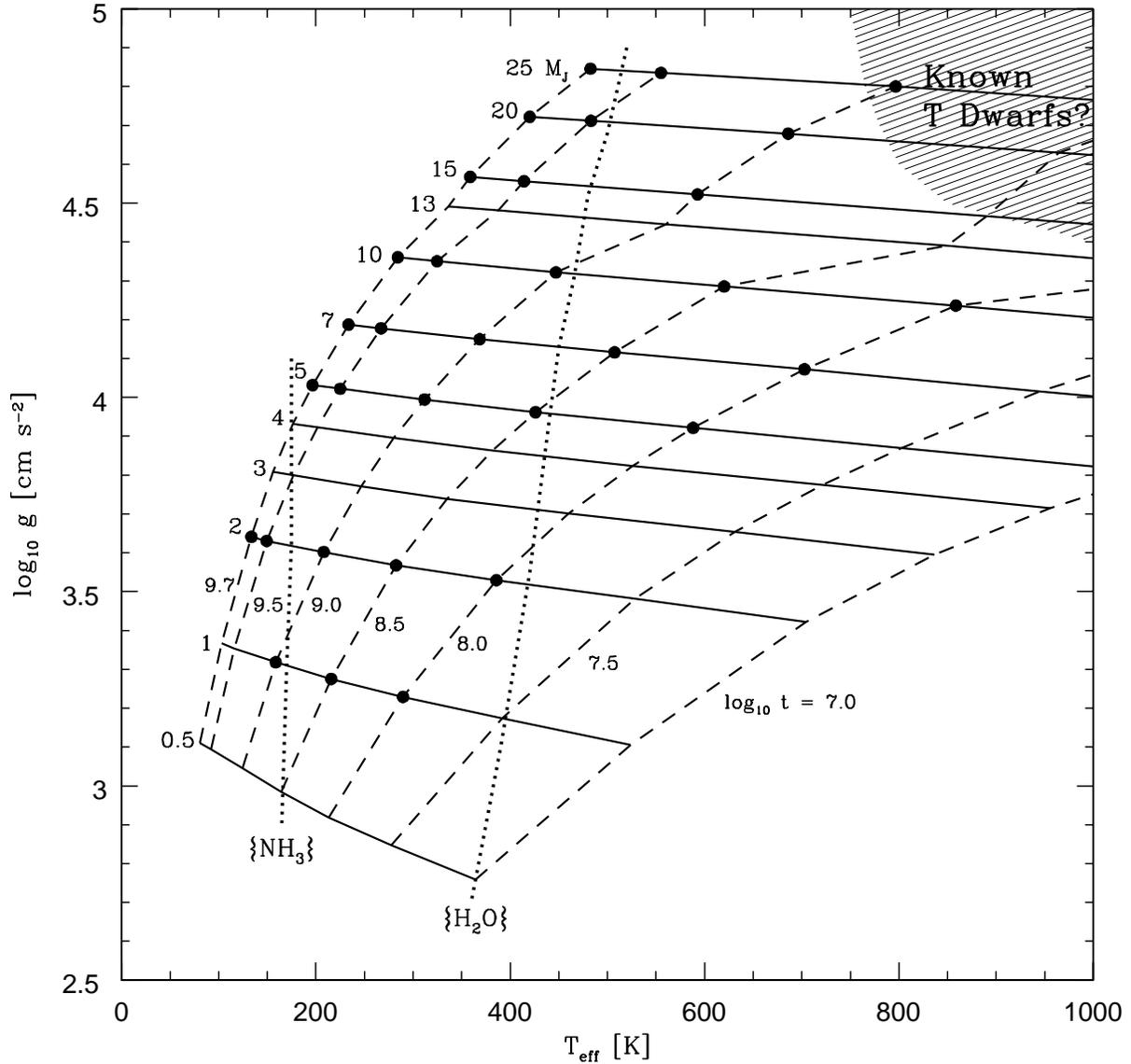}
\caption{Gravity versus effective temperature for a range
of brown dwarf masses (0.5 to 25 \mj) and ages (10 Myr
to 5 Gyr).  The solid curves depict evolutionary tracks,
while the dashed curves are isochrones.  The large dots denote  
the object parameters for which we calculate spectra and atmospheres 
for this paper.  The approximate condensation
curves for water and ammonia are plotted as dotted lines.
For objects to the left of these lines the corresponding condensate 
will form in the atmosphere.
As indicated, water is expected to condense in the atmospheres of a
sizable subset of these models, while ammonia is expected to condense
for only the lowest mass, oldest objects.
The hatched region in the top right identifies in approximate fashion 
where the currently known T dwarfs reside.   
As seen, they occupy only a small fraction 
of the depicted phase space.}
\label{fig:1}
\end{figure}

We have generated a new set of brown dwarf spectral models that incorporate  
state-of-the-art opacities and the effects of water clouds.  Our focus has
been on the low-\teff branch of the brown dwarf tree beyond the known T dwarfs.
To this end, we have investigated the \teff range from $\sim$800 K to $\sim$130 K
and the low-mass range from 25 to 1 \mj.  As Fig. \ref{fig:1} indicates, this is mostly
unexplored territory.  Our calculations have been done to provide a theoretical
foundation for the new brown dwarf studies that will be enabled by the launch of SIRTF and 
the eventual launch of JWST, as well as for the ongoing ground-based searches 
for the coolest substellar objects.  We provide spectra from $\sim$0.4 \mic to 30 \mic,
investigate the dependence on age and mass of the strengths of the H$_2$O, CH$_4$, and NH$_3$ 
molecular features, address the formation and effect of water clouds, and compare the calculated fluxes with 
the suggested sensitivities of the instruments on board SIRTF and JWST.  From the latter,
detection ranges can be derived, which for JWST can exceed a kiloparsec.  We find
that the blueward trend in near-infrared colors so characteristic of the T dwarfs 
stops near a \teff of 300-400 K and we identify a few natural physical transitions
in the low-\teff realm which might
justify the eventual designation of at least one new spectroscopic type after the T dwarfs.   
These include the formation of water clouds ($\sim$400-500 K), the strengthening
of ammonia bands, the eventual collapse in the optical, the shift in the position of the 
$M$ band peak, the turnaround of the $J-K$ color, the near disappearance of 
the strong Na-D and K I resonance lines ($\sim$500 K), and 
the increasing importance with decreasing \teff of the mid-IR
longward of 4 \mic. For these cooler objects, the mid-infrared assumes a new
and central importance and first MIPS and IRS on SIRTF, then MIRI on JWST,
are destined to play pivotal roles in their future characterization and study. 

Finally, the formation of ammonia clouds below $\sim$160 K
suggests yet another natural breakpoint, and a second new ``stellar" class.
Therefore, there are reasons to anticipate that perhaps two naturally defined, yet uncharted, spectral 
types reside beyond the T dwarfs at lower \teffs. 

The current filter set for JWST/NIRCam from 0.6 to 5.0 \mic is good, but not yet fully optimized
for cool brown dwarf detection.  Placing filters on the derived spectral peaks and troughs
(robustly defined by the water bands) would improve its already good performance
for substellar research.  In any case, our theoretical spectra are meant to bridge
the gap between the known T dwarfs and those cool, low-mass free-floating 
brown dwarfs with progressively more planetary features which may inhabit the
galaxy in interesting, but as yet unknown, numbers.

\acknowledgments
 
The authors thank Ivan Hubeny, Bill Hubbard,
John Milsom, Christopher Sharp, Jim Liebert, 
Curtis Cooper, and Jonathan Fortney
for fruitful conversations and help during the
course of this work, as well as
NASA for its financial support via grants NAG5-10760,
NAG5-10629, and NAG5-12459.

\begin{deluxetable}{cccc}
\tablewidth{8cm}
\tablecaption{Cool Brown Dwarf Model Grid}
\tablehead{
\colhead{$M/M_J$}  & \colhead{$\log_{10}  t\, (yr)$}  &  \colhead{$T_{\textrm{eff}}$ (K)}
& \colhead{$\log_{10} g$ (cm s$^{-2}$)}}
\startdata

   $1$ & $8.0$ & $290$ & $3.23$ \\
       & $8.5$ & $216$ & $3.27$ \\
       & $9.0$ & $159$ & $3.32$ \\
   $2$ & $8.0$ & $386$ & $3.53$ \\
       & $8.5$ & $283$ & $3.57$ \\
       & $9.0$ & $208$ & $3.60$ \\
       & $9.5$ & $149$ & $3.63$ \\
       & $9.7$ & $134$ & $3.64$ \\
   $5$ & $8.0$ & $588$ & $3.92$ \\
       & $8.5$ & $426$ & $3.96$ \\
       & $9.0$ & $312$ & $3.99$ \\
       & $9.5$ & $225$ & $4.02$ \\
       & $9.7$ & $197$ & $4.03$ \\
   $7$ & $8.0$ & $703$ & $4.07$ \\
       & $8.5$ & $507$ & $4.12$ \\
       & $9.0$ & $369$ & $4.15$ \\
       & $9.5$ & $267$ & $4.18$ \\
       & $9.7$ & $234$ & $4.19$ \\
  $10$ & $8.0$ & $859$ & $4.24$ \\
       & $8.5$ & $620$ & $4.29$ \\
       & $9.0$ & $447$ & $4.32$ \\
       & $9.5$ & $325$ & $4.35$ \\
       & $9.7$ & $284$ & $4.36$ \\
  $15$ & $9.0$ & $593$ & $4.52$ \\
       & $9.5$ & $414$ & $4.56$ \\
       & $9.7$ & $359$ & $4.57$ \\
  $20$ & $9.0$ & $686$ & $4.66$ \\
       & $9.5$ & $483$ & $4.71$ \\
       & $9.7$ & $421$ & $4.72$ \\
  $25$ & $9.0$ & $797$ & $4.80$ \\
       & $9.5$ & $555$ & $4.83$ \\
       & $9.7$ & $483$ & $4.85$ \\

\enddata
\end{deluxetable}

\newpage

\end{document}